\documentclass[final,5p,twocolumn]{elsarticle}

\newcommand{\stoptitle}{Search for light top squark pair production %
in final states with leptons and $b$-jets with the ATLAS detector in %
$\sqrt{s} = 7~\TeV$ proton--proton collisions}
\newcommand{\stopabstract}{The results of a search for pair production
of light top squarks are presented, using 4.7~\ifb of $\sqrt{s} =
7\TeV$ proton--proton collisions collected with the ATLAS detector at
the Large Hadron Collider. This search targets top squarks with masses
similar to, or lighter than, the top quark mass.  Final states
containing exclusively one or two leptons ($e, \mu$), large missing
transverse momentum, \ljets and \bjets are used to reconstruct the top
squark pair system. Event-based mass scale variables are used to separate
the signal from a large \ttbar background.  No excess over the
Standard Model expectations is found. The results are interpreted in
the framework of the Minimal Supersymmetric Standard Model, assuming
the top squark decays exclusively to a chargino and a \bquark, while
requiring different mass relationships between the Supersymmetric
particles in the decay chain. Light top squarks with masses between
123--167~\GeV are excluded for neutralino masses around 55~\GeV.}

\usepackage{atlasphysics}
\usepackage{xspace}
\usepackage{cancel}
\usepackage{preprintcover}  
\usepackage{hypernat}
\usepackage[breaklinks,colorlinks]{hyperref}
\hypersetup{linkcolor=blue,citecolor=blue,filecolor=black,urlcolor=blue} 

\usepackage[tight]{subfigure}

\usepackage{graphics}
\usepackage{graphicx}

\usepackage{amsmath}
\usepackage{amssymb}

\usepackage{lineno}

\biboptions{square,sort&compress}

\newcommand{\tone}   {\tilde{t}^{}_{1}}

\newcommand{\neut}  {\tilde{\chi}^{0}_{1}}

\def\bjets{$b$-jets\xspace}
\def\bjet{$b$-jet\xspace}

\def\ljets{light flavour jets\xspace}

\def\btagging{$b$-tagging\xspace}
\def\mistag{mis-tag\xspace}

\def\btagged{$b$-tagged\xspace}

\def\bquark{$b$-quark\xspace}

\def\mt{\ensuremath{m_\mathrm{T}}\xspace}

\def\Wjets{$W$+jets\xspace}
\def\Zjets{$Z$+jets\xspace}

\newcommand{\kt}{\ensuremath{k_{t}}\xspace}

\renewcommand{\ttbar} {\ensuremath{t\bar{t}}\xspace}
\renewcommand{\met} {\ensuremath{E_{\mathrm{T}}^{\mathrm{miss}}}\xspace}
\renewcommand{\pt} {\ensuremath{p_\mathrm{T}}\xspace}

\newcommand{\florida}{\ensuremath{\sqrt{s}_{\mathrm{min}}^{\mathrm{(sub)}}}\xspace}

\newcommand{\ptmiss}{\ensuremath{\mathbf{p}_\mathrm{T}^\mathrm{miss}}\xspace}
\newcommand{\lumi}{\ensuremath{4.7\pm0.2~\ifb}\xspace}
\newcommand{\hideit}[1]{{ }}

\newcommand{\ttv}[2]{{\tt {#1}{#2}}}

\newcommand{\alpgen}{{\tt ALPGEN}\xspace}
\newcommand{\powheg}{{\tt POWHEG}\xspace}

\newcommand{\herwig}{{\tt HERWIG}\xspace}
\newcommand{\herwigpp}{{\tt Herwig++}\xspace}
\newcommand{\pythia}{{\tt PYTHIA}\xspace}
\newcommand{\mcatnlo}{{\tt MC@NLO}\xspace}

\newcommand{\jimmy}{{\tt JIMMY}\xspace}
\newcommand{\cteqVER}[1]{{\tt CTEQ#1}\xspace}
\newcommand{\acermc}{{\tt AcerMC}\xspace}
\newcommand{\geant}{{\tt GEANT4}\xspace}
\newcommand{\madgraph}{{\tt MadGraph}\xspace}

\newcommand{\theendcap}{end-cap\xspace}
\newcommand{\theendcaps}{end-caps\xspace}
\newcommand{\mtoplep}{\ensuremath{m_t^{\mathrm{lep}}}\xspace}
\newcommand{\mtophad}{\ensuremath{m_t^{\mathrm{had}}}\xspace}
\newcommand{\mWlep}{\ensuremath{m_W^{\mathrm{lep}}}\xspace}
\newcommand{\mWhad}{\ensuremath{m_W^{\mathrm{had}}}\xspace}
\newcommand{\mll}{\ensuremath{m_{\lepton \lepton}}\xspace}

\newcommand{\mlljj}{\ensuremath{m_{\lepton \lepton jj}}\xspace}
\newcommand{\mjj}{\ensuremath{m_{jj}}\xspace}

\newcommand{\onelSR}{1LSR\xspace}

\newcommand{\twolSRone}{2LSR1\xspace}
\newcommand{\twolSRtwo}{2LSR2\xspace}

\newcommand{\cls}{CL$_s$\xspace}

\newcommand{\stopjournal}{Phys. Lett. B.}

\journal{\stopjournal}

\PreprintCoverPaperTitle{\stoptitle}  
\PreprintIdNumber{CERN-PH-EP-2012-207}  
\PreprintCoverAbstract{\stopabstract}  
\PreprintJournalName{\stopjournal}  

\begin{document}


\begin{frontmatter}



\title{\stoptitle}


\author{The ATLAS Collaboration}


\begin{abstract}
\stopabstract
\end{abstract}




\end{frontmatter}


\section{Introduction}
\label{sec:intro}
Supersymmetry (SUSY)~\cite{Miyazawa:1966,Ramond:1971gb,Golfand:1971iw,Neveu:1971rx,Neveu:1971iv,Gervais:1971ji,Volkov:1973ix,Wess:1973kz,Wess:1974tw}
is an extension of the Standard Model (SM) which naturally resolves the
hierarchy problem by introducing supersymmetric partners to the known
fermions and bosons. In the framework of a generic $R$-parity conserving
minimal supersymmetric extension of the SM (MSSM)~\cite{Fayet:1976et,Fayet:1977yc,Farrar:1978xj,Fayet:1979sa,Dimopoulos:1981zb},
SUSY particles are produced in pairs and the lightest supersymmetric
particle (LSP) is stable. In a large variety of models the LSP is the
lightest neutralino, $\tilde{\chi}^{0}_{1}$, which only interacts weakly. 
The scalar partners of right-handed and left-handed
quarks (squarks) can mix to form two mass eigenstates ($\squark_1, \squark_2$).
In particular, the lightest top squark (stop, $\tone$), could 
have a mass similar to, or lower than, the top quark ($t$) mass.

In this Letter, a search for direct stop pair production is presented
targeting these scenarios.  A SUSY particle mass hierarchy is assumed
such that $m_t \gtrsim m_{\tone} > m_{\chargino}$ and the stop decays
exclusively into a $b$-quark and a chargino ($\tone \rightarrow
\chargino b$). The chargino subsequently decays via a virtual or real
$W$ boson ($\tilde{\chi}^\pm_1 \rightarrow W^{(*)}
\tilde{\chi}^0_1 $).
The masses of all other supersymmetric particles, including the mass of $\tilde{t}_2$, are assumed to be 
above the~\TeV scale.
In the case where $m_{\tone} \sim m_t$,
direct stop pair production will lead to final states very similar to
SM \ttbar events, which form the dominant background. In the first
stage of the analysis the \ttbar system (including stop pairs) is
reconstructed from final states which contain exclusively one or two
leptons ($\lepton = e,\mu$), $b$-jets, light flavour jets, and large
missing transverse momentum. The use of event-based mass scale variables allows
discrimination between stop pairs and the \ttbar background.
The results are interpreted in three MSSM scenarios where stop and
neutralino masses are varied and different assumptions are made about
the chargino--neutralino mass difference: gaugino universality
($m_{\chargino} \simeq 2 \times m_{\neut}$); fixed chargino mass of
106~\GeV (above the present exclusion limit from LEP~\cite{lep2});
and fixed stop mass of 180~\GeV with variations of the chargino--neutralino
mass difference.
Previous results for direct production of top squark pairs 
in the same MSSM scenarios have been presented by the CDF~\cite{Aaltonen:2009sf}
and ATLAS Collaborations~\cite{ATLASlightstop}.

\section{The ATLAS Detector}
\label{sec:atlas}
The ATLAS detector is described in detail
elsewhere~\cite{DetectorPaper:2008}.  It comprises an inner detector (ID)
surrounded by a thin superconducting solenoid, a calorimeter
system and an extensive muon spectrometer embedded in
a toroidal magnetic field. The ID tracking system
consists of a silicon pixel detector, a silicon microstrip detector
(SCT), and a transition radiation tracker (TRT). It provides tracking
information for charged particles in a pseudorapidity\footnote{ATLAS
uses a right-handed coordinate system with its origin at the nominal
interaction point (IP) in the centre of the detector, with the
$z$-axis coinciding with the beam pipe axis. The $x$-axis points from
the IP to the centre of the LHC ring, and the $y$-axis points upwards.
Cylindrical coordinates $(r,\phi)$ are used in the transverse plane,
$\phi$ being the azimuthal angle around the beam pipe. The
pseudorapidity is defined in terms of the polar angle $\theta$ as
$\eta = -\ln \tan(\theta/2)$. The distance~$\Delta R$ in the
$\eta-\phi$ space is defined as $\Delta R =
\sqrt{(\Delta\eta)^2+(\Delta\phi)^2}$.}  range $|\eta|< 2.5$ and
allows efficient identification of jets originating from
$b$-hadron decays using impact parameter measurements to reconstruct
secondary decay vertices.  The ID is immersed in a $2$\,T axial
magnetic field and is surrounded by high-granularity liquid-argon
(LAr) sampling electromagnetic calorimeters.  An iron/scintillator
tile calorimeter provides hadronic energy measurements in the central
pseudorapidity range ($|\eta|<1.7$). In the forward regions
($1.5<|\eta|<4.9$), it is complemented by two end-cap calorimeters
using LAr as the active material and copper or tungsten as an
absorber.  The muon spectrometer (MS) surrounds the calorimeters and
consists of three large superconducting eight-coil toroids, a system
of tracking chambers, and detectors for triggering. The MS is segmented
into barrel ($|\eta| < 1.05$) and \theendcap regions ($1.05 < |\eta| < 2.7$).

\section{Simulated Event Samples}
\label{sec:sim}
Monte Carlo (MC) simulated event samples are used to develop and
validate the analysis procedure and to help evaluate the SM
backgrounds in the signal regions.
Production of top quark pairs is simulated with
\mcatnlo~\cite{Frixione:2002ik,Frixione:2003ei,Frixione:2005vw}, using a top quark mass of $172.5\GeV$
and the next-to-leading-order (NLO) parton distribution function
(PDF) set {\tt CT10}~\cite{Lai:2010vv}. Samples of $W$ and $Z$/$\gamma^*$ production,
with accompanying light- and heavy-flavour jets, and \ttbar with
additional \bjets ($\ttbar b \bar{b}$) are generated using
\alpgen~\cite{Mangano:2002ea}. 
Samples of $Z\ttbar$, $W\ttbar$ and $WW\ttbar$ are
generated with \madgraph~\cite{madgraph} interfaced to
\pythia~\cite{pythia}.
Diboson ($WW$, $WZ$, $ZZ$) production is generated with
\herwig~\cite{herwig}. Single top production is generated with
\mcatnlo for the $s$- and $t+W$-channels, and \acermc~\cite{Kersevan:2004yg}
for the $t$-channel.
Fragmentation and hadronisation modelling for the
\alpgen~and \mcatnlo~samples are performed by \herwig, using
\jimmy~\cite{Butterworth:1996zw} for the underlying event. \alpgen~and
\powheg~\cite{powheg}~samples are used to assess the systematic
uncertainties associated with the choice of generator for $t\bar{t}$
production, and \acermc samples are used to assess the
uncertainties associated with initial and final state radiation
(ISR/FSR).
The choice of PDF depends on the generator:
the \ttv{MRST}{2007} LO~\cite{ref:mrst2007lo} set is used with \herwig,
\cteqVER{6L1}~\cite{cteq6l} with \alpgen.
The background predictions are normalised to the theoretical cross
sections, including higher-order QCD corrections when available, as
detailed in Ref.~\cite{ATLAS:2012ah}.

Direct stop pair production samples are generated using \ttv{PYTHIA}{6}
and \herwigpp~\cite{Bahr:2008pv}.
Polarisation effects due to the choice of left- and right-handed scalar
top mixing were found to have a negligible impact on the analysis.
Signal cross sections are calculated to NLO in the
strong coupling constant, adding the resummation of soft gluon
emission at next-to-leading-logarithmic accuracy
(NLO + NLL)~\cite{Beenakker:1997ut,Beenakker:2010nq,Beenakker:2011fu}.

All MC samples are produced using a detector
simulation~\cite{atlassimulation} based on \geant~\cite{geant4}. MC
samples are re-weighted such that the number of additional
proton--proton interactions per bunch crossing (pile-up) agrees with
that observed in data.

\section{Event Reconstruction and Preselection}
\label{sec:objreco}
Electron candidates are reconstructed from energy clusters in the
electromagnetic calorimeters matched to a track in the ID.  They are
required to have momentum in the transverse plane (\pt) $\pT >
20\GeV$, $|\eta| < 2.47$ and to pass the ``medium'' shower shape and
track selection criteria of Ref.~\cite{Aad:2011mk}.

Muons are reconstructed using an algorithm~\cite{muonefficiency} that
combines information from the ID and MS.  Candidate muons are required
to have $\pT > 10\GeV$, $|\eta| < 2.4$, and be reconstructed with
sufficient numbers of hits in the pixel, SCT and TRT detectors.  In
order to reject muons originating from cosmic rays, events containing
muon candidates with a closest approach distance greater than
1~mm to the primary vertex in the $z$ direction, or a transverse impact
parameter greater than 0.2~mm, are rejected.  The primary vertex
itself is defined as the vertex with the highest summed track $\pT^2$.

Jet candidates are reconstructed using the anti-\kt jet clustering
algorithm~\cite{Cacciari:2008gp} with a radius parameter of
$R=0.4$. The measured jet energy is corrected for inhomogeneities and
for the non-compensating nature of the calorimeter using \pt{} and
$\eta$ dependent correction factors based on MC simulation validated
with extensive test-beam and collision-data
studies. 
Furthermore, the reconstructed jet is
modified such that its direction points to the primary vertex, and
events containing jets likely to have arisen from detector noise or
cosmic rays are rejected~\cite{Aad:2011he}. Only jet candidates with
corrected transverse momenta $\pT > 20\GeV$ and $|\eta|<4.5$ are retained.

Following their reconstruction, candidate jets and leptons may point
to the same energy deposits in the calorimeter. These overlaps are
resolved by first discarding any jet candidate within $\Delta R = 0.2$
of an electron. Then, any electron or muon candidate remaining within
$\Delta R = 0.4$ of any surviving jet is also discarded.

The two-dimensional missing transverse
momentum vector, \ptmiss, and its magnitude \met, are computed from the
negative of the vector sum of the \pT~of the reconstructed electrons, muons
and jets, and all energy clusters with $|\eta| < 4.9$ not
associated with such objects.

Electrons must additionally pass the
``tight'' electron criteria of Ref.~\cite{Aad:2011mk},
and be isolated such
that the scalar~\pT~sum of tracks within a cone of $\Delta R = 0.2$
around the electron candidate (not including the electron track) must
be less than 10\% of the electron
\pT. Muons must also be isolated such that the \pT~sum of tracks (not including
the muon track) within $\Delta R = 0.2$ is less than 1.8~\GeV. Jets are
further required to lie within $|\eta| < 2.5$ and must have
more than 75\% of \pt-weighted ID tracks associated to the primary vertex.
This reduces the presence of jets arising from uncorrelated soft collisions
(pile-up) and discards jets without reconstructed tracks.

A \btagging algorithm~\cite{ATLAS:BtagCalib} is used to identify
jets containing a $b$-hadron decay. The algorithm uses a
multivariate technique based on the properties of the secondary
vertex, of tracks within the jet, and of the jet itself. The nominal
$b$-tagging efficiency, determined from $t\bar{t}$ MC events, is on
average 60\%, with a misidentification, or mis-tag, rate for $c$-quark
(light-quark/gluon) jets of 10\% (1\%).


\section{Signal Region Definitions}
\label{sec:evsel}
The data are selected with a three-level trigger system.
The events used in this search satisfied single-lepton trigger
requirements that varied with the data-taking period.
The tightest
electron trigger has an efficiency of $\sim$97\% for electrons with
$\pT > 25\GeV$. The muon trigger reaches an efficiency
plateau of $\sim$90\% in the \theendcaps for muons with $\pT > 20\GeV$. 
The equivalent efficiency in the barrel region is $\sim$75\%
due to a lower geometrical acceptance for the muon trigger chambers
in this region.
Collision events are selected by requiring
at least one reconstructed vertex with at least five associated tracks with $\pt > 400\MeV$,
consistent with the beam spot position. Two signal regions are
defined containing either exclusively one or two charged leptons
($\lepton = e, \mu$) in the final state, referred to hereafter as the
1- and 2-lepton channels respectively.
A total integrated luminosity of \lumi is used, following the
beam, detector and data-quality requirements as described in Refs.~\cite{Aad:2011dr,lumicorrect}.

In the 1-lepton channel, events are required to contain the minimum
number of objects expected from the
$\ttbar \rightarrow WbW\bar{b} \rightarrow q\bar{q}^{\prime}b \lepton \bar{\nu} \bar{b}$
decay.
Exactly one lepton is required, which must have $\pT >
25\GeV$ ($20\GeV$) for the electron (muon) channel and fulfil the
trigger requirements. Events with an additional electron (muon) with
$\pT > 20\GeV$ ($10\GeV$) are rejected to ensure no events are classified
as belonging to both 1- and 2-lepton channels.
A minimum of four jets are required in the
event, at least two of which must pass the \btagging requirements and at least two
must fail them. Events are required to have a missing transverse momentum of
$\met > 40\GeV$. Background from multi-jet processes, in which jets
are misidentified as leptons, is rejected by requiring that the
transverse mass of the lepton-\met system,
$\mt = \sqrt{2\pT^{\lepton{}}\met - 2\mathbf{p}_\mathrm{T}^{\lepton{}} \ldotp \ptmiss}$,
is larger than $30\GeV$.

\begin{figure*}[hbt!!]
\centering
\subfigure[]{\label{fig:plot_mTopHad}\includegraphics[width=0.49\textwidth]{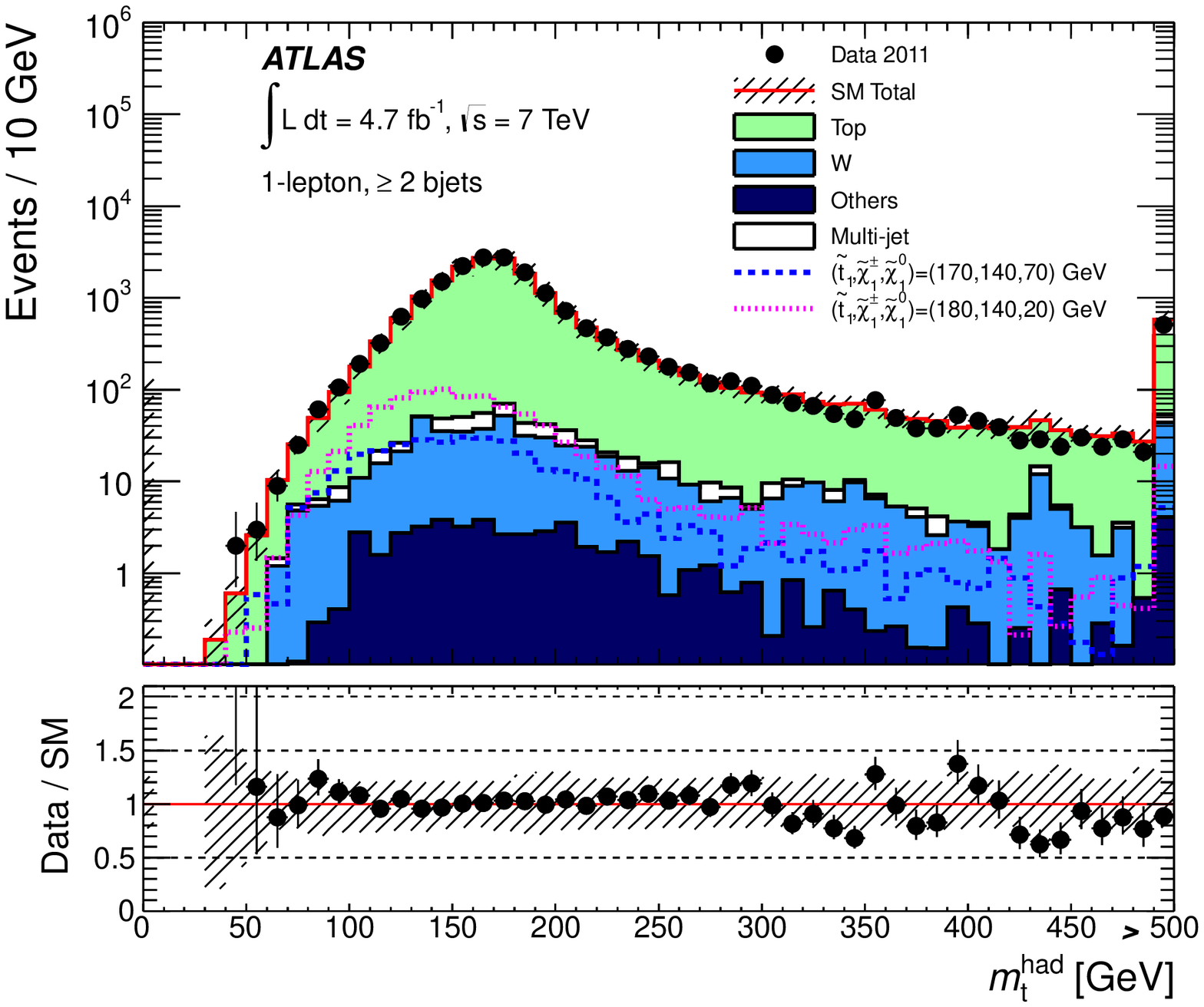}}
\subfigure[]{\label{fig:onelepflorida}\includegraphics[width=0.49\textwidth]{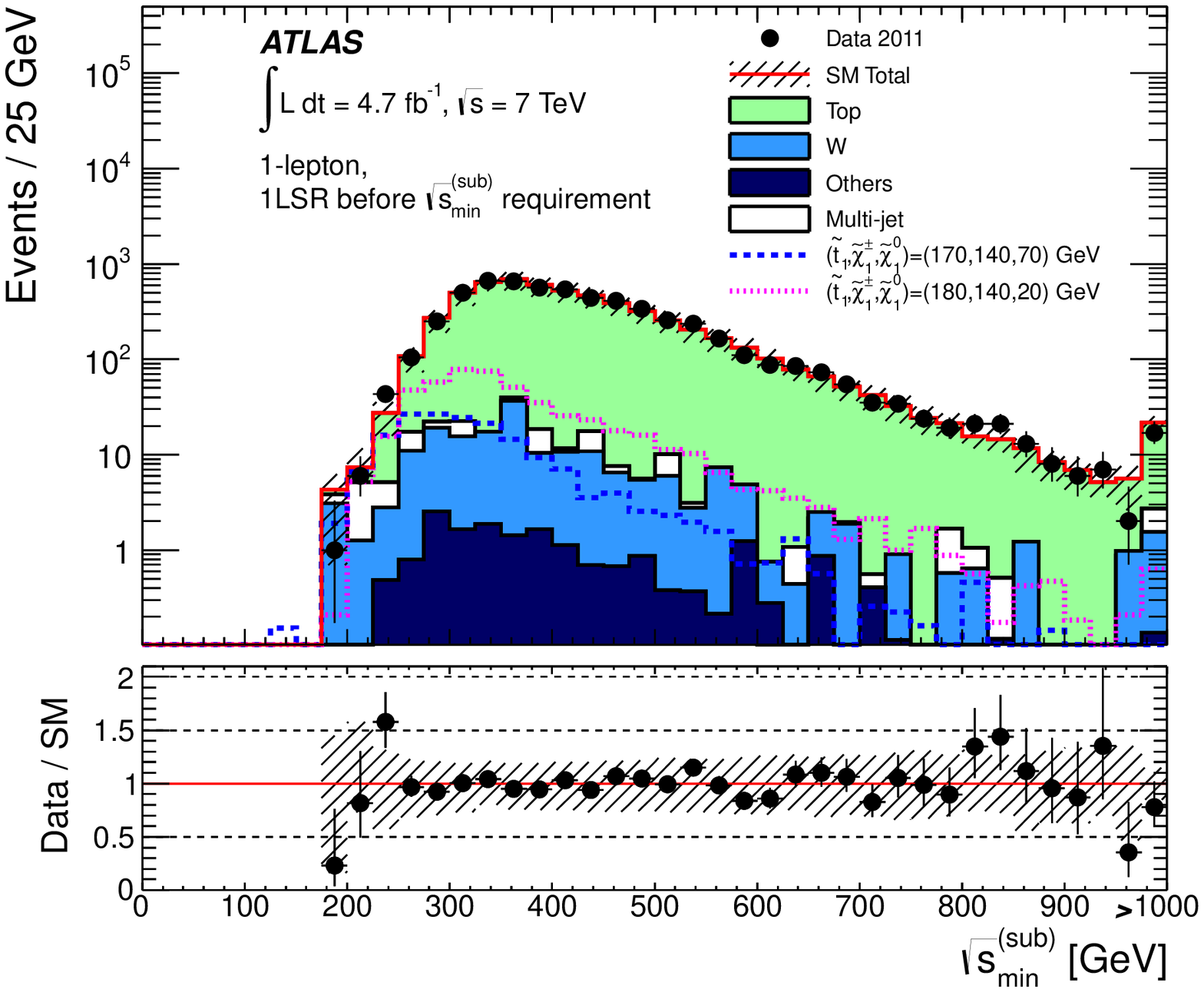}}

\subfigure[]{\label{fig:twolepmll}\includegraphics[width=0.49\textwidth]{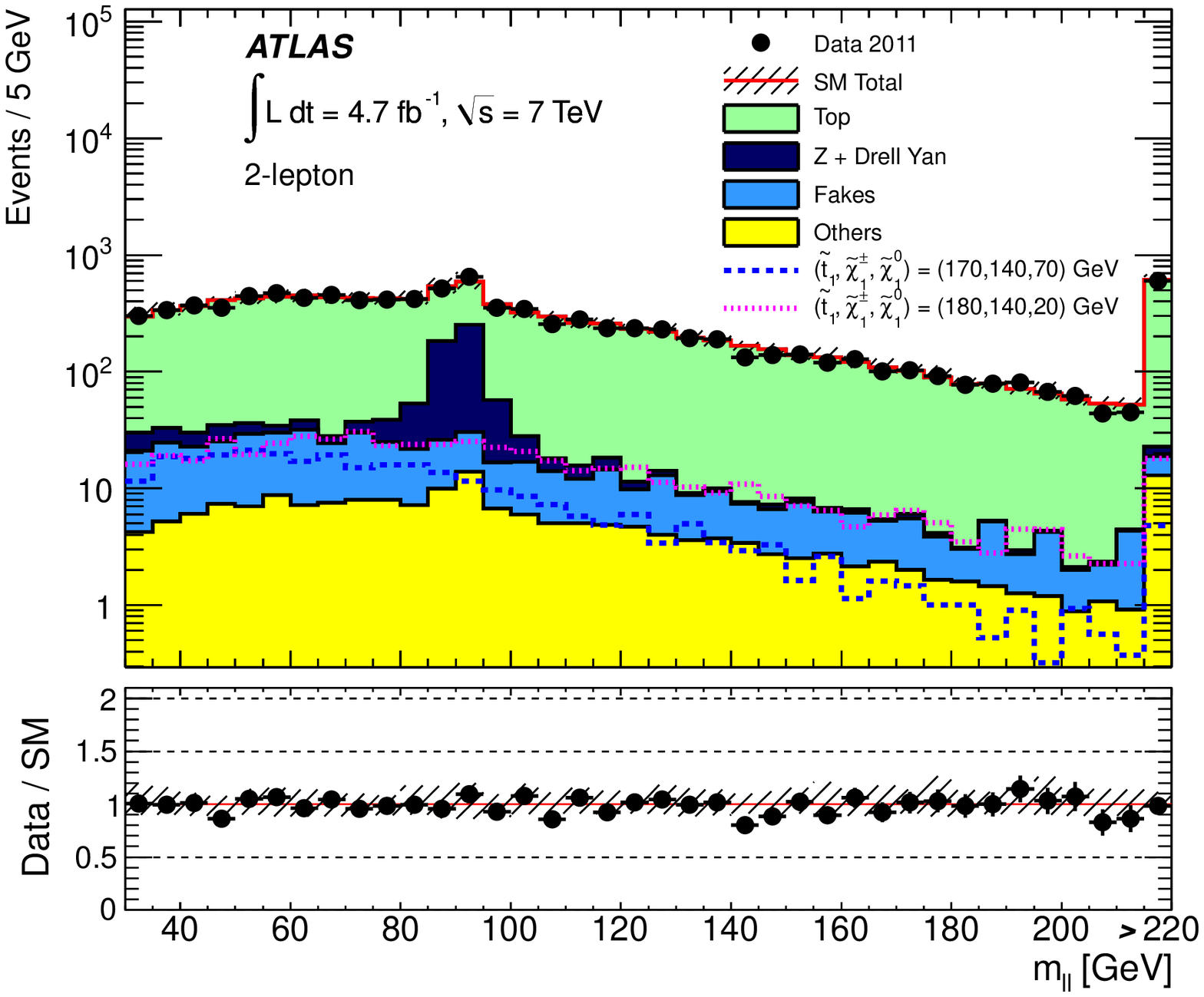}}
\subfigure[]{\label{fig:twolepflorida}\includegraphics[width=0.49\textwidth]{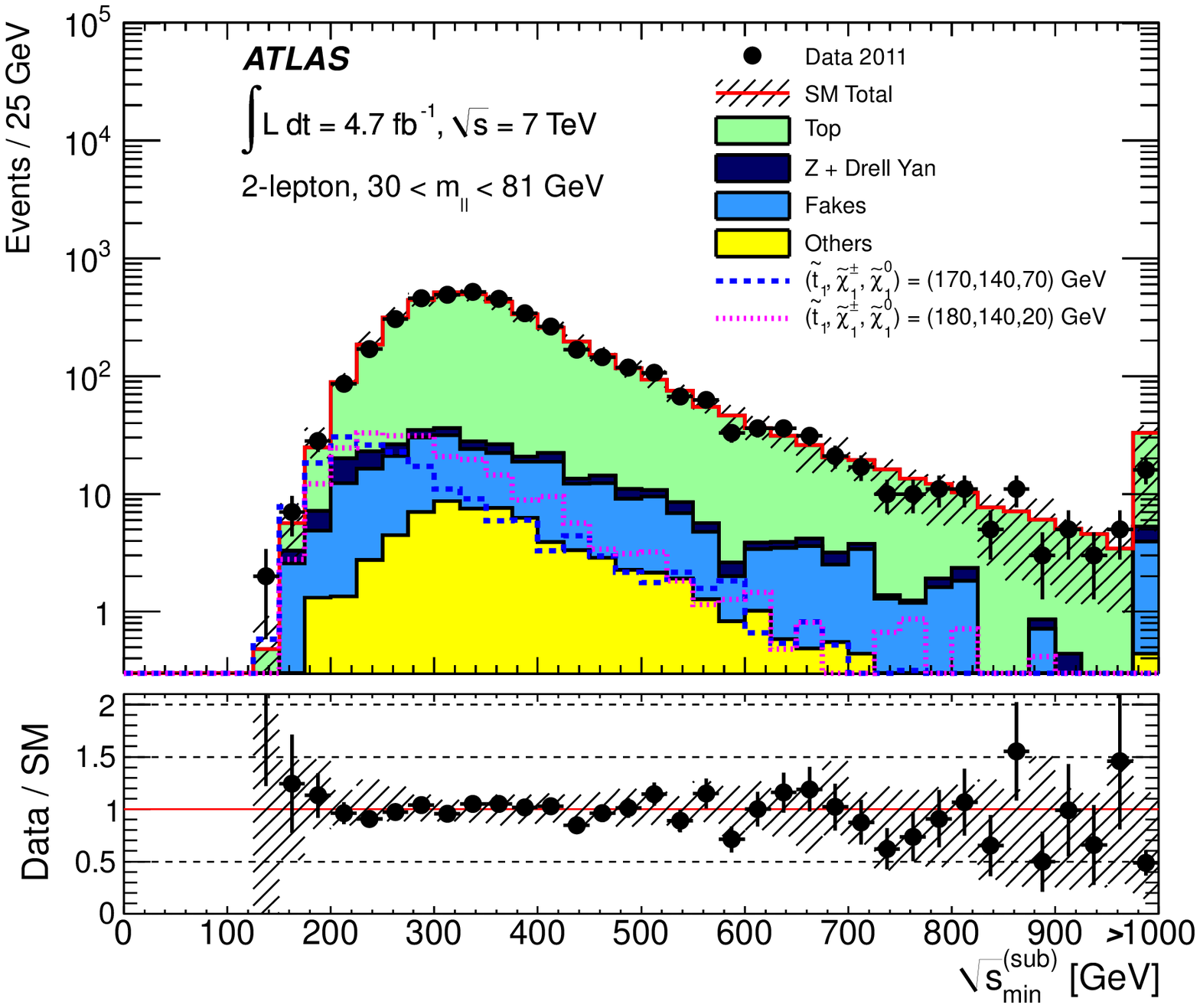}}

\caption{The 1-lepton channel \mtophad distribution after all
requirements except those on \mtophad and \florida~(a), and the
\florida distribution after all requirements except that on
\florida~(b). For the 2-lepton channel, the \mll distribution is shown
after all requirements except those on \mll and \florida{}~(c), and
the \florida distribution, before the requirements on \florida
itself~(d). The last bin in each histogram contains 
the integral of all events with values greater than the upper axis bound. 
The hatched bands display the total uncertainties on the
background expectation and
the dashed lines show the expected distributions for two signal models.
The bottom panels show the ratio of data to
the expected background (points) and the total uncertainty on the
background (hatched area). }
\label{fig:prettyplots}
\end{figure*}

The invariant mass of the hadronic top decay products (\mtophad) is
used as an additional discriminating variable. In scenarios where the
stop is lighter than the top, \mtophad will tend to be lower than for
background \ttbar processes, as illustrated in Fig.~\ref{fig:plot_mTopHad}.
Since there is an ambiguity as to which \bjet arises from the hadronic
top decay (and additional ambiguities at higher jet multiplicities),
the hadronic decay products are tagged using the following algorithm:
for every possible combination of light and \btagged jets in the
event, the invariant masses \mWhad (of two light jets, \mjj),
\mWlep (assuming that the lepton and \met arise from the $W \rightarrow
\lepton \nu$ decay), \mtophad and \mtoplep (the leptonic top mass)
are calculated. A \ttbar estimator of $P_{\mathrm{tot}} = P(\mWhad)
P(\mWlep) P(\mtoplep) P(\mtophad)$ is assigned to this combination,
where $P(m)$ is related to the probability for reconstructing a particle of mass
$m$, assuming a Gaussian probability density function with mean values
taken from Ref.~\cite{pdg2012} and widths from MC simulation. The
combination which maximises $P_{\mathrm{tot}}$ is used to assign one
\bjet and two
\ljets (one \bjet, the lepton and \met) as arising from the hadronic
(leptonic) decay of the top quark.

Events are then required to have $\mtophad < \hat{\mu} -
0.5\hat{\sigma}$, where $\hat{\mu}$ and $\hat{\sigma}$ are the mean
and width respectively of a Gaussian fit to the \mtophad distribution
in a 40~\GeV window around the top mass.  This approach is taken to
reduce some of the systematic uncertainties affecting the shape of this
distribution, as detailed in Section~\ref{sec:background}.

In the 2-lepton channel, the following requirements are imposed to
ensure that the event contains the required number of objects
consistent with the 
$\ttbar \rightarrow WbW\bar{b} \rightarrow \lepton \nu b \lepton \bar{\nu} \bar{b}$ decay.
Exactly
two oppositely-charged leptons are required to pass the selection
described in Section~\ref{sec:objreco}.  For same-flavour pairs, the
highest \pt lepton is required to have $\pT > 25\GeV$ ($20\GeV$) for
electrons (muons). In the case of different-flavour pairs, either the
electron must have $\pT > 25\GeV$ or the muon $\pT >20\GeV$.  At least
two jets are required in the event, of which the two with highest \pt
are assumed to originate from the \ttbar process. At least one of
these two jets is required to be \btagged.
The event is required to fulfil  $\met > 40\GeV$
and the invariant mass of the two leptons $(\mll)$ must
satisfy $30 < \mll < 81\GeV$ to increase the discrimination against the background, as illustrated in Fig.~\ref{fig:twolepmll}.

In order to distinguish between stop and top pair production the mass
scale subsystem variable~\florida{}~\cite{florida2011} is employed.
Conceptually, the variable is constructed by dividing an event's topology
into a subsystem comprising both the visible and invisible
particles originating from the hard-scattering process of interest 
and a set of other
visible particles labelled as coming from other, ``upstream'', processes such as
the underlying event or ISR.  With these definitions, the minimum
invariant mass compatible with the subsystem is:

\begin{equation}\label{eq:florida}
\begin{split}
\florida = & \left\{\left(\sqrt{m^2_{(\mathrm{sub})}+{\pt}^2_{(\mathrm{sub})}}
+ \sqrt{ \left(m^{\mathrm{miss}}\right)^2 + \left( {\met}\right)^2 }\right)^2 \right. \\
- & \left. \left(\mathbf{\pt}_{(\mathrm{sub})}+\ptmiss \right)^2 \right\}^{\frac{1}{2}},
\end{split}
\end{equation}
where $m_{(\mathrm{sub})}$ and $p_{\mathrm{T}(\mathrm{sub})}$ are the invariant mass and the
transverse momentum of the visible subsystem particles.
The variable $m^{\mathrm{miss}}$ is the scalar sum of the invisible particle
masses in the event. The final term in
Eq.~\ref{eq:florida} is a two-dimensional vector sum representing the boost
correction in the transverse plane caused by upstream processes.
In this analysis \florida is calculated making the hypothesis that each
event arises from \ttbar production, with the invisible subsystem
comprising one or two neutrinos, and therefore $m^{\mathrm{miss}} = 0$
in Eq.~\ref{eq:florida}.  With this assumption, the reconstructed \florida
distribution for
\ttbar background events is expected to 
peak at around $m_{\ttbar} = 2m_t \simeq 345\GeV$.  On the other hand,
stop pair production will peak at lower values if the mass difference
between the stop and the neutralino is 
less than the top mass, as illustrated in Figs.~\ref{fig:onelepflorida}
and~\ref{fig:twolepflorida}.
Signal events are therefore selected by imposing an upper limit on
\florida.

In the 1-lepton channel, the visible subsystem comprises the single
lepton, two \ljets and two \bjets. In events where combinatorial
ambiguities arise, the subsystem objects are chosen which give the
highest estimator in the algorithm described above.
In the 2-lepton channel, the two leptons
and the two leading jets are used. In both channels, the upper limit on
\florida has been chosen to maximise the expected
signal efficiency with respect to background rejection, across a range
of scenarios described in Section~\ref{sec:intro}.
In the 1-lepton channel, the optimal requirement is $\florida < 250\GeV$,
defining a signal region referred to hereafter as \onelSR.
%
%
In the 2-lepton channel two signal regions are defined, the first
requiring $\florida < 225\GeV$ (\twolSRone). The invariant mass of the
two leptons and two jets (\mlljj) was also found be a useful
discriminating variable.  Imposing $\mlljj < 140~\GeV$ in combination
with $\florida < 235\GeV$ was found to give the optimal performance and
defines a second signal region (\twolSRtwo).

\setlength{\subfigcapskip}{-0.5cm}
\setlength{\subfigbottomskip}{0.2cm}
\setlength{\subfigtopskip}{0cm}

\begin{table*}[!!tbp]
  \centering
    \begin{tabular}{l|l|l|l}
    \hline
    \hline
            & \multicolumn{3}{c}{Number of events} \\
    \hline
    Process & \onelSR &\twolSRone  & \twolSRtwo \\
    \hline
    Top    & $24 \pm 3 \pm 5$ & $89 \pm 6 \pm 10 $ & $36 \pm 2 \pm 5$\\
    \Wjets & $6 \pm 1 \pm 2$ & --- & --- \\
    \Zjets & $0.5 \pm 0.3 \pm 0.3$ &  $11 \pm 4 \pm 3  $ & $3 \pm 1 \pm 1 $\\
    Fake leptons  & $7 \pm 1 \pm 2$ &  $12 \pm  5 \pm 11$ & $6 \pm 4 \pm 4$\\
    Others & $0.3 \pm 0.1 \pm 0.1$ &  $2.7 \pm 0.9 \pm 0.7 $ & $0.9 \pm 0.2 \pm 0.5 $\\
    \hline
    Total SM & $38 \pm 3 \pm 7 $ & $115 \pm 8 \pm 15$ & $46 \pm 4 \pm 7$\\
    \hline
    Data  & 50 & 123 & 47\\    
    \hline
    $m_{\tone}=170\GeV, m_{\neut}=70\GeV$ &
               $26 \pm 2 \pm 6$ & $57 \pm 3 \pm 6$ & $36 \pm 2 \pm 4$ \\
    $m_{\tone}=180\GeV, m_{\neut}=20\GeV$ &
               $20 \pm 2 \pm 4$ & $41 \pm 3 \pm 5$ & $27 \pm 2 \pm 3$ \\
    \hline
            & \multicolumn{3}{c}{95\% CL upper limits} \\
    \hline
    $\sigma_{\mathrm{vis}}$ (expected) [fb] & 4.2 & 9.3 & 4.6 \\
    $\sigma_{\mathrm{vis}}$ (observed) [fb] & 6.1 & 11  & 5.2 \\
    \hline
    \hline
    \end{tabular}%
  \caption{Predicted and observed number of events in all signal
regions together with their statistical and systematic uncertainties.
No values are shown for the \Wjets contributions in the
2-lepton signal regions as these are included in the fake contributions. The expected number
of events for two signal scenarios, both with a chargino mass of 140~\GeV, are also shown.
The observed and expected upper limits at 95\% confidence level on
$\sigma_{\mathrm{vis}} = \sigma \times A \times \epsilon$ are also
given.}
  \label{tab:results}%
\end{table*}%

\section{Background Estimation}
\label{sec:background}
The dominant SM background process in the 1-lepton (2-lepton) channel
arises from single-lepton (dilepton) \ttbar decays, comprising 60\%
(80\%) of the total background. The second most significant background
in the 1-lepton (2-lepton) channel arises from $W$ ($Z/\gamma^*$)
production in association with jets from heavy flavour quarks.
For both channels, similar methods are used to estimate these backgrounds.
For each channel and background process a control region is
defined that is rich in the background of interest. The region is
kinematically similar to the signal region but distinct from it,
such that the signal and control regions have no events in common.
For a control region containing $N^{\textrm{obs}}_{\textrm{CR}}$ observed events
(corrected for the contamination from other backgrounds), the number of
events in the signal region is calculated as $N_{\textrm{SR}} = 
N^{\textrm{obs}}_{\textrm{CR}} \times (N^{\textrm{MC}}_{\textrm{SR}}  /
N^{\textrm{MC}}_{\textrm{CR}} )$, where $N^{\textrm{MC}}_{\textrm{SR}}$
and $N^{\textrm{MC}}_{\textrm{CR}}$ are the MC-based estimates in the signal
and control regions respectively.  The advantage of this method is that
many systematic uncertainties partially cancel.

In the 1-lepton channel, the \ttbar background is determined with a
control region defined identically to the signal region, except
that $\hat{\mu} - 0.5\hat{\sigma} < \mtophad < \hat{\mu} + 0.5
\hat{\sigma}$ and $\florida<320\GeV$, corresponding to a \ttbar purity of 93\%.
 The definition of a control region using these fitted parameters
reduces the systematic uncertainties related to the jet energy scale
and resolution.  A high-purity $W$+\bjets control region is more
difficult to define due to the kinematic similarity with \ttbar
events, which have a higher fiducial cross section. A control region
can, however, be defined with 38\% purity for $W$+\bjets events by
requiring $\mtophad > 250\GeV$ and that the invariant mass of the two
\bjets is less than 50~\GeV.  As the \ttbar contamination in this
region is still relatively high (60\%), the $W$+\bjets and \ttbar
contributions are determined by scaling their contributions
simultaneously such that the total number of events matches the data
in both control regions.

In the 2-lepton channel, the \ttbar background (including dileptonic
$Wt$ decays) is determined using a control region identical to the
signal region except that $\mll > 101\GeV$ and $\florida < 325\GeV$,
with 94\% purity of \ttbar events. The $Z+$jets background,
with $Z$ decaying to any of the three lepton flavours, is determined in a region
requiring two same-flavour leptons, $81 < \mll < 101\GeV$ and
$\florida < 225\GeV$, with a $Z$ purity of 90\%. 

The contribution to the background from events where a jet
is misidentified as a lepton, or where a lepton from a $b$-
or $c$-hadron decay is selected (referred to as ``fake'' lepton background),
is estimated using a data-driven technique in both channels~\cite{Aad:2011mk,Aad:2011cwa}.
The probability of such a misidentification is estimated by relaxing the
electron and muon identification criteria to obtain control samples dominated
by multi-jet production. In the 1-lepton channel, the main
contribution is from multi-jet events. In the 2-lepton channel,
the dominant contribution is from processes containing one real and
one fake lepton, such as $W$+jet or single-lepton
\ttbar decays. The contribution from events containing two fake
leptons was found to be negligible.

Other less significant processes in the 1-lepton channel include
$Z/\gamma^*$+jets and single top quark production. Diboson and
$\ttbar+X$ $(X=W,Z,WW,b\bar{b})$ production give a minor contribution to both
channels. The contribution to the total background from these
processes (referred to as ``Others'' in the following and in Fig.~\ref{fig:prettyplots})
is 2.5\% (2\%) in the 1-lepton (2-lepton) channel, and is
taken directly from the MC predictions.

\section{Systematic Uncertainties}
\label{sec:syst}
The effect of the jet energy scale (JES) uncertainty on the final
event yield is calculated by shifting the \pt of all jets up and down
by \pt and $\eta$ dependent factors, which are 5--3\% for jets with
\pt of 20--60~\GeV. Repeating the analysis with these \pt shifts
applied to the MC simulation leads to variations on the final
background estimate of 6--10\% depending on the signal region. The
uncertainty due to the jet energy resolution (JER) is calculated by
smearing the \pt of each jet by factors depending on the jet \pt and
$\eta$. The smearing on a single jet is typically around 10\%, and
results in an overall uncertainty of 1--10\%. Systematic uncertainties
in the lepton identification efficiency amount to 1\%.  The
uncertainty on the \met due to the energy scale of the clusters in the
calorimeter not associated with jets and electrons is evaluated using
the method described in Ref.~\cite{Aad:2012re}, extended to include
pile-up uncertainties. The effect is up to 9\% on the total background
estimate depending on the signal region.  The uncertainty due to \btagging is
evaluated by varying the
\btagging efficiency and \mistag rates within the uncertainties of the
measured
values~\cite{ATLAS:bTagEfficiency,ATLAS:MisTagbRate,ATLAS:cTagCalib},
giving an effect of 1\% in all signal regions.  The uncertainty
associated with pile-up re-weighting is evaluated by varying the
number of interactions per bunch-crossing by 10\%.  The overall effect
on the predicted background yield is at most 3\%.

Uncertainties related to the overall normalisation of the top
background are reduced compared to estimates based purely on MC
simulation by employing the method described in
Sec.~\ref{sec:background}. Residual uncertainties related to
the shape of the predicted kinematic distributions are described
in the following.
Theoretical uncertainties on the \ttbar background due to the choice
of generator are evaluated by comparing event yields from \mcatnlo to
those from \powheg with the same parton shower model (\herwig). The
parton shower uncertainties are then calculated by comparing samples
generated with the \herwig and \pythia parton shower models, with the
same generator (\powheg). The uncertainty due to ISR/FSR is assessed
using \acermc samples with variations of \pythia parameters related to
the ISR branching phase-space and the FSR low-\pt cutoff. These
variations are chosen to produce jet activity in \ttbar events that is
consistent with the data~\cite{ATLAS:2012ISR, ATLAS:2012FSR}. The
total uncertainty on the \ttbar estimate due to these effects amounts
to 10--15\%. Uncertainties due to the PDF choice and errors are found
to be negligible.

In the 1-lepton channel, the theoretical uncertainty in the $W$
estimate due to variations of the factorisation, renormalisation and
matching scales is found to be 15\%. Similar uncertainties on the
$Z/\gamma^*$ contribution in the 2-lepton channel are 9\% (2\%) in
\twolSRone (\twolSRtwo).

Uncertainties on the data-driven background from fake leptons
arise from the lepton fake rate determination and from the definition
of the fake-enriched control regions. The effect is between 45--84\%
of the fake contribution.

Theoretical uncertainties on the stop pair production cross section are
taken from an envelope of predictions which use different PDF sets and
factorisation and renormalisation scales, as described in
Ref.~\cite{Kramer:2012bx}. 
Signal uncertainties on JES (10--30\%), JER (1--30\%) and \btagging
(5--10\%) vary depending on the sparticle masses and the signal channel
considered. They are treated as fully correlated with their respective
background uncertainties. Finally, the luminosity uncertainty is 3.9\%.

\setlength{\subfigcapskip}{-0.55cm}
\setlength{\subfigbottomskip}{0cm}
\setlength{\subfigtopskip}{0cm}

\begin{figure}[!htp]
\centering
\subfigure[]{\label{fig:limitgs}\includegraphics[width=0.41\textwidth]{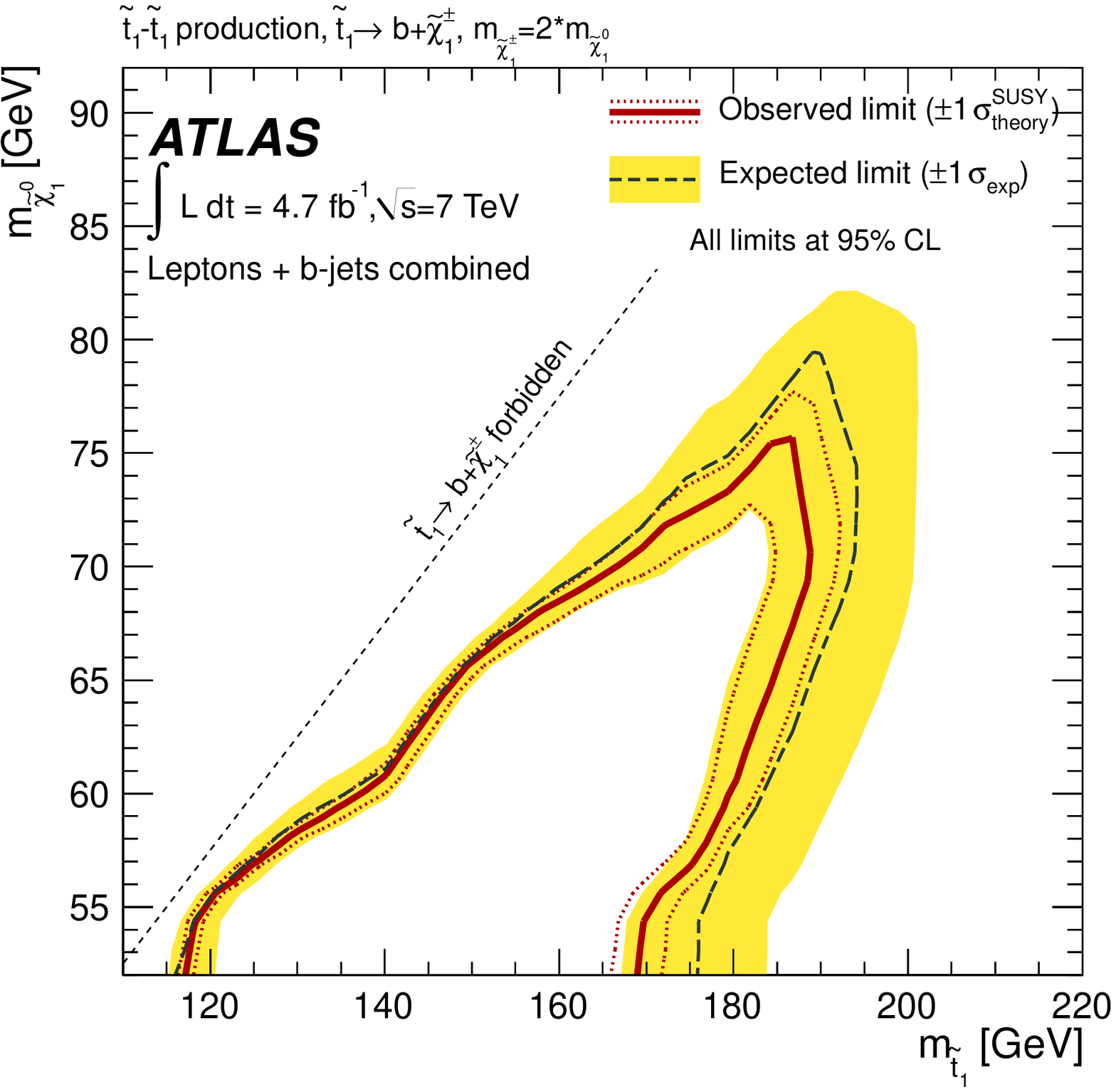}}
\subfigure[]{\label{fig:limitstop180}\includegraphics[width=0.41\textwidth]{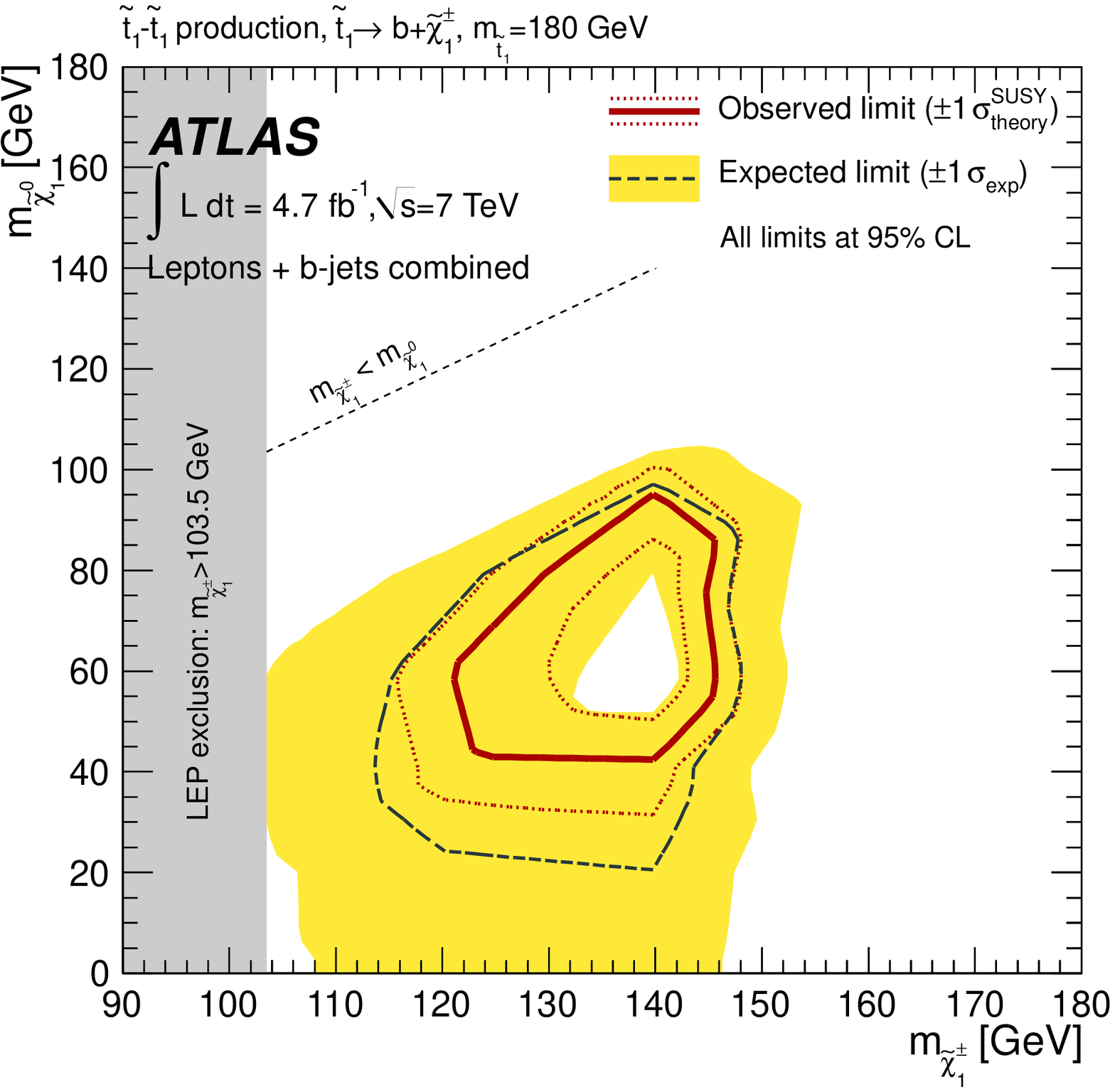}}
\subfigure[]{\label{fig:limitC106}\includegraphics[width=0.41\textwidth]{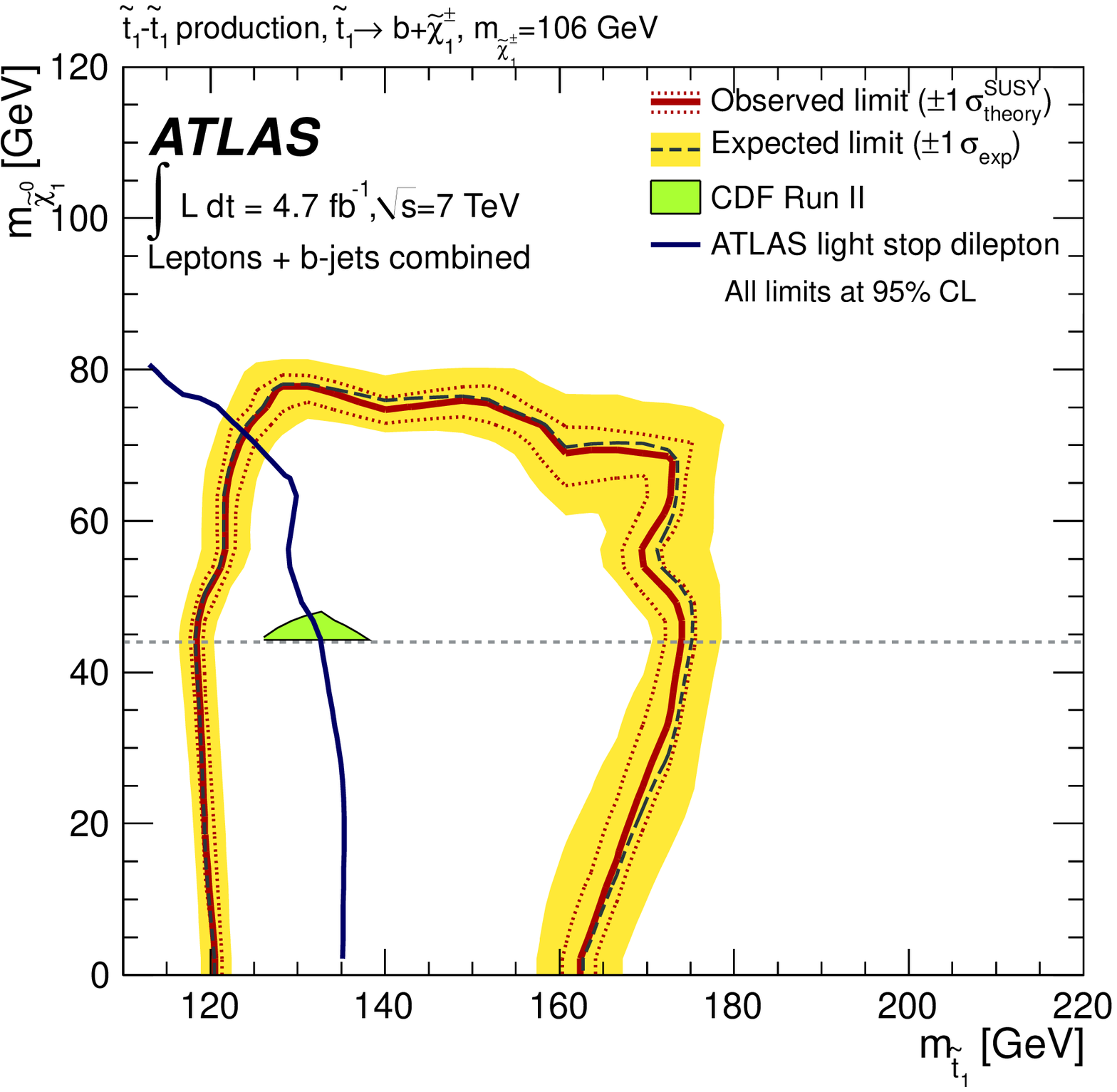}}
\caption{
Exclusion limits at 95\% CL for the scenarios described in the text.
The dashed (solid) lines show the expected (observed)
limits, including all uncertainties except for the
theoretical signal cross section uncertainty (PDF and scale). The
bands around the expected limits show the $\pm1\sigma$
results. The lines around the observed limits represent the results
obtained when moving the nominal signal cross section up or down by
the $\pm1\sigma$ theoretical uncertainty. In~(c),
results are compared to previous limits from the Tevatron~\cite{Aaltonen:2009sf},
where the lowest neutralino mass considered was $44\GeV$ (dotted line),
and from ATLAS~\cite{ATLASlightstop}.
}
\label{fig:limitstopneut}
\end{figure}

\section{Results and Interpretation}
\label{sec:results}
Table~\ref{tab:results} shows the observed number of events in data and the
SM predictions for the signal regions of the 1- and 2-lepton channels. 
In all SRs, the data are in good agreement with the SM
expectations.

The results are translated into 95\% confidence-level (CL) upper
limits on contributions from new physics using the \cls
prescription~\cite{Read:2002hq} with a profile log-likelihood ratio as
a test statistic~\cite{likelihood}, where the parameter to describe
the non-SM signal strength is constrained to be non-negative in the
fit.  As shown in Table~\ref{tab:results}, the three signal regions
are used to set limits on the visible cross section of the new physics
models, $\sigma_{\mathrm{vis}}=\sigma \times A \times \epsilon$, where
$\sigma$ is the total production cross section for any non-SM signal,
$A$ is the acceptance defined by the fraction of events passing the
geometric and kinematic selections at particle level, and $\epsilon$
is the detector reconstruction, identification and trigger efficiency.
Results are interpreted in the MSSM scenarios described in
Section~\ref{sec:intro}.
In order to maximise the sensitivity of the analysis, results from the
1- and 2-lepton channels are combined using the following method: for
each signal point, the 2-lepton signal region
(\twolSRone or \twolSRtwo) which yields the lowest expected \cls value is
chosen. This region is then statistically combined with the \onelSR by
multiplying the respective likelihood functions. Correlated systematic
uncertainties are treated as common between the two channels, and a
common signal strength parameter $\mu$ is applied.
The effect of signal contamination in the control regions (typically
5--10\% depending on the signal point) is also considered. In the
gaugino universality scenario, shown in Fig.~\ref{fig:limitgs}, stop
masses between 120--167~\GeV are excluded for $m_{\neut}=55\GeV$.  The
sensitivity of the search is also evaluated for a stop mass of
180~\GeV in the chargino--neutralino mass plane, as shown in
Fig.~\ref{fig:limitstop180}.  In such a scenario, where the stop-top
mass difference is small, a region around $m_{\neut}=70\GeV$,
$m_{\chargino}=140\GeV$ is still excluded.  The scenario with
$m_{\chargino}=106\GeV$ is shown in Fig.~\ref{fig:limitC106}, where
stop masses are excluded between 123--167~\GeV for $m_{\neut}=55\GeV$.
Neutralino masses of 70~\GeV are excluded for $125 < m_{\tone} <
155\GeV$.

\section{Conclusions}
\label{sec:conclusions}
A search has been performed for top squarks with masses near to, or less
than, the top quark mass. 
Good agreement is observed between data and
the SM predictions in all channels.
The results allow limits to be set on the stop mass,
assuming that
$\tone \rightarrow \chargino b$ is the only allowed decay mode,
followed by $\tilde{\chi}^\pm_1 \rightarrow W^{(*)}
\tilde{\chi}^0_1 $.
For scenarios in which $m_{\chargino} \simeq 2 \times m_{\neut}$,
stop masses between 120--167~\GeV are excluded for $m_{\neut}=55\GeV$.
For a fixed stop mass of 180~\GeV, a region around $m_{\neut}=70\GeV$,
$m_{\chargino}=140\GeV$ is excluded.
In the scenario where $m_{\chargino}=106\GeV$, neutralino masses
of 70~\GeV are excluded for $125 < m_{\tone} < 155\GeV$,
significantly extending previous limits in such scenarios.


\section{Acknowledgements}

We thank CERN for the very successful operation of the LHC, as well as the
support staff from our institutions without whom ATLAS could not be
operated efficiently.

We acknowledge the support of ANPCyT, Argentina; YerPhI, Armenia; ARC,
Australia; BMWF, Austria; ANAS, Azerbaijan; SSTC, Belarus; CNPq and FAPESP,
Brazil; NSERC, NRC and CFI, Canada; CERN; CONICYT, Chile; CAS, MOST and NSFC,
China; COLCIENCIAS, Colombia; MSMT CR, MPO CR and VSC CR, Czech Republic;
DNRF, DNSRC and Lundbeck Foundation, Denmark; EPLANET and ERC, European Union;
IN2P3-CNRS, CEA-DSM/IRFU, France; GNAS, Georgia; BMBF, DFG, HGF, MPG and AvH
Foundation, Germany; GSRT, Greece; ISF, MINERVA, GIF, DIP and Benoziyo Center,
Israel; INFN, Italy; MEXT and JSPS, Japan; CNRST, Morocco; FOM and NWO,
Netherlands; RCN, Norway; MNiSW, Poland; GRICES and FCT, Portugal; MERYS
(MECTS), Romania; MES of Russia and ROSATOM, Russian Federation; JINR; MSTD,
Serbia; MSSR, Slovakia; ARRS and MVZT, Slovenia; DST/NRF, South Africa;
MICINN, Spain; SRC and Wallenberg Foundation, Sweden; SER, SNSF and Cantons of
Bern and Geneva, Switzerland; NSC, Taiwan; TAEK, Turkey; STFC, the Royal
Society and Leverhulme Trust, United Kingdom; DOE and NSF, United States of
America.

The crucial computing support from all WLCG partners is acknowledged
gratefully, in particular from CERN and the ATLAS Tier-1 facilities at
TRIUMF (Canada), NDGF (Denmark, Norway, Sweden), CC-IN2P3 (France),
KIT/GridKA (Germany), INFN-CNAF (Italy), NL-T1 (Netherlands), PIC (Spain),
ASGC (Taiwan), RAL (UK) and BNL (USA) and in the Tier-2 facilities
worldwide.

\bibliographystyle{apsrev4-1}
\bibliography{bjetEarlySummer}

\onecolumn
\clearpage
\begin{flushleft}
{\Large The ATLAS Collaboration}

\bigskip

G.~Aad$^{\rm 48}$,
T.~Abajyan$^{\rm 21}$,
B.~Abbott$^{\rm 111}$,
J.~Abdallah$^{\rm 12}$,
S.~Abdel~Khalek$^{\rm 115}$,
A.A.~Abdelalim$^{\rm 49}$,
O.~Abdinov$^{\rm 11}$,
R.~Aben$^{\rm 105}$,
B.~Abi$^{\rm 112}$,
M.~Abolins$^{\rm 88}$
% ...
\bigskip
\\
% ...
$^{11}$ Institute of Physics, Azerbaijan Academy of Sciences, Baku, Azerbaijan\\
$^{12}$ Institut de F{\'\i}sica d'Altes Energies and Departament de F{\'\i}sica de la Universitat Aut{\`o}noma de Barcelona and ICREA, Barcelona, Spain\\
% ...
$^{21}$ Physikalisches Institut, University of Bonn, Bonn, Germany\\
% ...
$^{48}$ Fakult{\"a}t f{\"u}r Mathematik und Physik, Albert-Ludwigs-Universit{\"a}t, Freiburg, Germany\\
$^{49}$ Section de Physique, Universit{\'e} de Gen{\`e}ve, Geneva, Switzerland\\
% ...
$^{88}$ Department of Physics and Astronomy, Michigan State University, East Lansing MI, United States of America\\
% ...
$^{105}$ Nikhef National Institute for Subatomic Physics and University of Amsterdam, Amsterdam, Netherlands\\
% ...
$^{111}$ Homer L. Dodge Department of Physics and Astronomy, University of Oklahoma, Norman OK, United States of America\\
$^{112}$ Department of Physics, Oklahoma State University, Stillwater OK, United States of America\\
% ...
$^{115}$ LAL, Universit{\'e} Paris-Sud and CNRS/IN2P3, Orsay, France\\
% ...
\end{flushleft}


\clearpage

\end{document}
